\documentstyle[12pt]{article}
\begin{document}
\centerline{\bf \large ON DESCRIPTION OF THE DIRECT NUCLEON DECAY}
\centerline{\bf \large OF GIANT RESONANCES} 
\bigskip
\centerline{\bf G.A.Chekomazov, S.E.Muraviev,  M.H.Urin} 
\medskip
\centerline{\it Moscow Engineering Physics Institute, 115409, Moscow, Russia}
\newcommand{\bld}[1]{\lefteqn{#1}\hspace{0.05em}\lefteqn{#1}%
\hspace{0.05em}{\raisebox{-.05ex}{$#1$}}}
\medskip

1. The direct nucleon decay of giant resonances (GR) in intermediate
and heavy mass nuclei is now the subject of intensive experimental and
theoretical studies. At the moment there is a body of experimental
data on the GR nucleon decay to certain (single-hole) states of a
product nucleus. These data are: (i) the partial proton escape widths
of isobaric analog resonance (IAR) (see, e. g., \cite{1,2}
and refs. therein);
(ii) the partial cross sections of the photonucleon
reactions accompanied by excitation of the giant dipole resonance
(GDR) (see, e.g., \cite{3});
(iii) the rather contradictory data on the
partial neutron escape widths of the isoscalar
monopole giant resonance (GMR) in $^{208}Pb$ \cite{4,5}; 
(iv) the branching ratios for the direct neutron 
decay  of the GDR and region above in $^{208}Pb$ \cite{6}; 
(v) the partial proton escape widths of the 
Gamow-Teller resonance (GTR) in $^{208}Bi$ \cite{7}.
The appearance of new experimental data on the direct nucleon decay
of various GR is expected in the near future (see, e.g., these Proceedings).

The theoretical description of the direct nucleon decay of GR is a serious
test for models of nuclear structure and nuclear reactions,
because single-particle, collective, and many-particle
aspects of nuclear motion are combined in this phenomenon.
Modern theoretical approaches to
the mentioned description are based on the random
phase approximation with consideration for the single-particle
continuum (on the continuum-RPA) [2,4,8-17].
In these refs. the energies, strengths, and partial escape
widths of collective states of the particle-hole (p-h) type (these states
are called below the doorway states (dws), or the gross-structure components
of the considered GR) have been 
calculated on the basis of different versions of the continuum-RPA
with the use of
different forms of the nuclear mean field and p-h
interaction. Attempts to consider the GR nucleon decay
together with explicit consideration for the coupling of the
dws to 2p-2h configurations have been undertaken in  \cite{11,14}. 

In the present talk the somewhat extended version of the
approach developed in  \cite{15,17} is briefly outlined.
Applications of the approach to description of the direct 
nucleon decay of a number of GR in the $^{208}Pb$ parent nucleus
and a comparison of calculation results with relevant
experimental data are given.

2. The main features of the approach are the following.
(i) As input data for the RPA calculations we use
the phenomenological mean field of the Woods-Saxon type and the 
phenomenological p-h interaction.
This latter is chosen
in different forms 
to clarify the dependence of the calculation
results on the choice of the p-h interaction.
For the description of IAR and GTR we realize the partial 
selfconsistency (see, e. g., \cite{2}), which is a result of the
approximate isospin-symmetry of the nuclear Hamiltonian. 
(ii) We use the exact version of the 
continuum-RPA for calculating the amplitudes
of those partial $(e,N_c)$- and 
inclusive reactions,  which are studied experimentally 
($e=N_c, \gamma,...$ means the entrance reaction-channel: nucleon, radiative,
...;  
$c$ is the set of quantum numbers for the nucleon decay-channel).
(iii) In the case of GR with not-too-large excitation energy 
it is possible to use the
Breit-Wigner parametrization of the 
reaction amplitudes calculated within the continuum-RPA for evaluating the
dws parameters: the excitation energy, the entrance width (or corresponding
strength), the partial escape widths.
(iv) We consider the doorway-state coupling to many-particle configurations
phenomenologically on the average over the
energy using a reasonable statistical assumption and
the fact that there are no common channels for the nucleon decay
of the dws and many-particle states \cite{15}.
This consideration allows us to describe
the energy-averaged reaction-amplitudes 
by means of an independent spreading (and energy shift) of each dws
resonance in the RPA reaction-amplitudes.
(v) The adequate consideration
of the penetrability of the potential barrier for the
escaping nucleon requires the use of (a) the experimental
energies of the nucleon decay-channels instead of calculated ones 
\cite{16}; (b) the penetrability averaged over the GR region.
(vi) The possibility to describe the direct nucleon decay of GR
in terms of the escape widths depends on the structure of the
GR strength function (or of the inclusive reaction cross-section)
calculated within the continuum-RPA. (a) When only one dws
corresponds to the GR, the energy-averaged reaction
cross-sections in the GR region look like a Breit-Wigner resonance. 
In this case the reaction cross-sections can be described in terms of the
partial nucleon escape widths $\Gamma_{c}$
(or of the widths $\langle \Gamma_{c}\rangle $ averaged over the GR region), 
which are coincident with those calculated within the continuum-RPA
and are independent of both the dws coupling to many-particle 
configurations and the entrance channel~\cite{15,17}. For this reason
the entrance channel can be chosen for convenience of 
calculations. We usually use a nucleon channel as entrance one
\cite{15,16}. 
(b) In the case of some essential gross-structure it is necessary
to calculate the energy-averaged cross-sections
of the $(eN_c)$- reactions $\sigma_{eN_c}$ (or the branching
ratios $b_{c}^{(e)}=\sigma_{eN_c} / \sigma_{e}$, where 
$\sigma_{e}$ is the energy-averaged inclusive reaction cross-section)
and to compare them with experimental data directly. Only in
the case, when the root-mean-squared energy dispersion of the dws
is less than the spreading width 
of the dws $\Gamma^\downarrow$, the calculated cross section  
$\sigma_{eN_c}$ can be approximately represented in the Breit-Wigner form,
and the branching ratios are equal to 
$b_{c}^{(e)}=\Gamma_{c}^{(e)} / \Gamma^\downarrow$, where
$\Gamma_{c}^{(e)}$ are effective partial widths of the GR.
These widths are determined by the amplitudes of the entrance
and escape widths of the dws and, therefore,
are dependent on the entrance channel \cite{17}. 

3. To illustrate the abilities of the approach outlined briefly
above let us turn to description of the direct nucleon decay
for a number of GR in the $^{208}Pb$ parent nucleus. The
parametrization and parameters of the phenomenological mean 
field used in the calculations are given in detail in \cite{15}.
The choice of the phenomenological p-h interaction is given
below for each considered GR.

\noindent
\underline{\bf Isobaric analog resonance.} The theoretical description of the
direct proton decay of the IAR is closely related to the approximate
conservation of the isospin symmetry in nuclei (see, e. g., \cite{2,18}
and refs. therein). If this symmetry were exact, all nucleon decays of 
the IAR would be forbidden. Therefore, these decays are determined
by the mechanism and strength of the isospin-symmetry violation in
nuclei. 
To take the violation into account most correctly 
within the shell-model, the isovector part
of the nuclear mean field (the symmetry potential) should be 
consistent with the neutron-excess density 
via the isovector part
of the p-h interaction $F_\tau(x,x')=
\bld{\tau}\bld{\tau}' F_\tau({\bf r},{\bf r'})$,
where $x$ is the set of nucleon coordinates including spin and isospin ones
 (see, e. g., \cite{2}).
In calculations we use the Landau-Migdal and separable interactions: \\
\\
$F_\tau({\bf r},{\bf r'})=F_\tau\delta({\bf r}-{\bf r'}),$ \hfill  (1a) \\
\\
$F_\tau({\bf r},{\bf r'})=K_\tau f_{ws}(r)f_{ws}(r'),$ \hfill (1b) \\
\\
where $f_{ws}(r)$ is the Woods-Saxon function.
The strength parameters $F_\tau=f'\cdot 300$\ MeV$\cdot\mbox{fm}^3$,
$ f'=0.95$,
and $K_\tau=0.16$ \ MeV are found  by using the mentioned
selfconsistency condition and the phenomenological expressions for
the symmetry potential and the neutron-excess density.
Because the Fermi strength function calculated with the use of interactions
(1) has no gross-structure in the IAR region, we calculate the
S-matrix of the proton-nucleus scattering and evaluate the 
partial proton widths of the IAR in the $^{208}Bi$ by means
of the Breit-Wigner parametrization of the S-matrix using the
experimental energies of the decay channels. The calculated widths 
are given in Table 1 together with experimental widths deduced
from both the resonance reactions (see refs. in \cite{18})
and the direct reactions \cite{7}. Because the total width of the
IAR is small, there is no need to use
the penetrability averaged over the resonance.
As expected, the calculated widths $\Gamma_c$ are dependent on the choice
of the p-h interaction only slightly, because these widths are mainly
determined by the mean Coulomb field \cite{2,18}.      \\
\leftline{\bf Table 1}
\leftline{\it Proton escape widths (in keV) of IAR and GTR in $^{208}Bi$}
\begin{tabular}{cccccccccc}
\hline
final & \multicolumn {4} { c } {IAR} && 
\multicolumn {4} { c } {GTR} \\
\cline {2-5}
\cline {7-10}
state & \multicolumn {2} { c } {$\Gamma_c$(exp)}& 
\multicolumn {2} { c } {$\Gamma_c$(calc)}& &
 {$\Gamma_c$(exp)}& 
 {$\Gamma_c$(calc)}& 
 \multicolumn {2} { c } {$\langle \Gamma_c \rangle$ (calc)} \\
of $^{207}Pb$ & \cite{18}&\cite{7}&int.&int.&&\cite{7}&\cite{15}&
int.&int. \\
&&& (1a)& (1b) &&&& (2a) & (2b) \\
\hline
$1/2^-$ & 51.6$\pm$1.7 & 51.4$\pm$5.6 & 61 & 63 && 48.9$\pm$9.3 
& 33 & 43 &173 \\
$5/2^-$ & 24$\pm$4 & incl.in $p_{3/2}$ & 27 & 31 && 
incl.in $p_{3/2}$& 18 & 48 &164  \\
$3/2^-$ & 54$\pm$7 & 79.4$\pm$9.4 &79 & 86 & & 84.9$\pm$13.1 & 21 & 35 & 218 \\
$13/2^+$ & & &0.18&0.17 & & 6.9$\pm$7.7&0.04&0.78&5.2 \\
$7/2^-$ & 5.0$\pm$0.5 & 3.5$\pm$1.6 &8.5&8.7& &13.1$\pm$6.2&0.26&8.9&78  \\
\hline
$\Gamma_{tot}^\uparrow$ &134.6&134.3&176&189& &153.8&72&136&638 \\
\hline
\end{tabular}
\medskip

\noindent
\underline{\bf Gamow-Teller resonance}. The partial proton
escape widths $\Gamma_{c}$ of the GTR in $^{208}Bi$ 
have been calculated in \cite{15} and \cite{16}
with the use of the Landau-Migdal p-h interaction
(respectively, the calculated and experimental energies of the proton channels
have been used in these refs). In the present work we use
the spin-isospin part of the p-h interaction   
$F_{\sigma \tau}(x,x')=
(\bld{\sigma} \bld{\sigma}')(\bld{\tau} \bld{\tau}') F_{\sigma\tau}
({\bf r},{\bf r'})$ in the following forms:\\
\\
$F_{\sigma\tau}({\bf r},{\bf r'})=F_{\sigma\tau}
\delta({\bf r}-{\bf r'}),$ \hfill (2a)\\
\\
$F_{\sigma\tau}({\bf r},{\bf r'})=
K_{\sigma\tau} f_{ws}(r)f_{ws}(r').$ \hfill (2b)\\
\\
The strength parameters in eqs. (2) 
$F_{\sigma\tau}=g' \cdot 300$\ MeV$\cdot \mbox{fm}^3,\ g'=0.76$ and 
$K_{\sigma\tau}=0.11$\ MeV
are found so that the energy of the main maximum
of the GT strength function calculated within the continuum-RPA
would be equal to the experimental GTR energy.
Notice in this connection that in the present work
we use the determination of $F_{\sigma\tau}$
 which is different from that has been used in
the original edition of the monograph \cite{19} and in \cite{15,16}.
This difference may not lead to some misunderstanding of our previous
results.

Because the GT strength function calculated with the use of interactions
(2) has no gross-structure in the
GTR region, we evaluate the widths $\Gamma_c$ by the same way
as in the case of the IAR. However, 
because the observable total width of the GTR is rather large
($\Gamma_{tot}=$3.75 \ MeV~\cite{7}), we recalculate the
widths $\Gamma_{c}$ to $\langle \Gamma_{c}\rangle $
using the proton penetrabilities averaged over the resonance.
The calculated averaged widths $\langle \Gamma_{c}\rangle $ are 
systematically lager
than nonaveraged ones (compare the results given in the seventh
and eighth columns of Table 1). The comparison of two last
columns of Table 1 shows that the use of the separable
interaction (2b) leads to the marked overestimation of the partial escape
widths of the GTR.

\noindent
\underline{\bf Isoscalar monopole resonance}. The partial neutron escape widths
$\Gamma_{c}$ of the GMR in $^{208}Pb$
have been calculated in \cite{12,15} with 
the use of the Landau-Migdal
p-h interaction 
and calculated energies of the neutron decay channels.
The calculated
widths have been markedly overestimated as compared with
the relevant experimental widths \cite{4,5}.
In the present work we use the isoscalar part of the p-h interaction
$F(x,x')=F({\bf r},{\bf r'})$  in
two forms: \\
\\
$F({\bf r},{\bf r'})=F(r)\delta({\bf r}-{\bf r'})$ \hfill (3a)\\
\\
$F({\bf r},{\bf r'})=
-K_0 (df_{ws}(r) / dr)
(df_{ws}(r') / dr')$ \hfill (3b) \\
\\
The radial-dependent strength $F(r)$ 
of the Landau-Migdal
p-h interaction (3a) is a two-parametric function with a sharp change at 
the nuclear surface (see, e. g., \cite{15}).
The strength parameter $K_0=5.8$\ MeV$\cdot\mbox{fm}^2$ in eq. (3b) 
is found so that the 
calculated energy of the GMR is consistent with experimental one.
Because the monopole strength function calculated with
the use of interactions (3) has no gross-structure 
in the GMR region, 
we evaluate the partial neutron escape widths  
$\langle \Gamma_{c}\rangle$ by the same way, which is used 
in the case of the GTR. 
Calculated results are given in Table 2 together with the relevant
experimental data \cite{4,5}. \\
\leftline{\bf Table 2}
\leftline{\it Neutron escape widths (in keV) of GMR and GDR in $^{208}Pb$}
\begin{tabular}{cccccccccc}
\hline
final & \multicolumn {4} { c } {GMR} & &
\multicolumn {4} { c } {GDR} \\
\cline {2-5}
\cline {7-10}
state & \multicolumn {2} { c } {$\Gamma_c$(exp)}& 
\multicolumn {2} { c } {$\langle \Gamma_c\rangle$(calc)}& &
 {$\Gamma_c^{(\gamma)}$}& 
 {$\Gamma_c^{(n)}$}& 
 {$\Gamma_c^{(\gamma)}$}&
 {$\Gamma_c^{(n)}$}  \\
of $^{207}Pb$ & \cite{5}&\cite{4}&int.(3a)&int.(3b)& &
\multicolumn {2} { c } {int.(1a)}& 
\multicolumn {2} { c } {int.(4)} \\
\hline
$1/2^-$ & 0 & 140$\pm$35 & 76 & 49 && 42 & 75 & 1 & 23 \\
$5/2^-$ & $<$35 & 70$\pm$15 & 430 & 259 && 173 & 411 & 27 & 62 \\
$3/2^-$ & 75$\pm$40 & 50$\pm$10 & 156 & 96 && 71 & 103 & 17 & 42 \\
$13/2^+$ & 75$\pm$35 & incl.in $3/2^-$ &11 & 1.5 && 9 & 188 & 6 & 247 \\
$7/2^-$ & $<$140$\pm$30 & 165$\pm$40 & 323 & 169 && 7 & 7 & 31 & 38 \\
$9/2^-$ &&&&&& 8 & 11 & 8 & 11 \\
\hline
$\Gamma_{tot}^\uparrow$ & 325 & 425 & 996 & 574 && 310 & & 90 & \\
\hline
\end{tabular}
\medskip

\noindent
The use of the separable interaction
(3b) somewhat decreases the calculated widths. Nevertheless,
they are also overestimated as compared to the experimental widths.
It can be expected that the use of the separable interaction
as a superposition of surface and volume parts could allow
us to describe the experimental widths adequately. These calculations
are in progress now. 

\noindent
\underline{\bf Giant dipole resonance}.
The GDR is an example of the giant resonance exhibiting 
some essential gross-structure
within the continuum-RPA (see, e. g.,  \cite{9,15}). 
To describe the direct nucleon decay of such a GR it is
necessary to calculate the cross sections 
$\sigma_{eN_c}$ and $\sigma_{e}$ of the reactions which are
studied experimentally. An example
of the calculation is given in  \cite{17}, where the radiative
channel has been considered as entrance one, so that $\sigma_{eN_c}$ 
and $\sigma_{e}$ are the cross sections of the
partial $(\gamma N_c)$-reaction and photoabsorption, respectively.
The satisfactory description of the experimental 
$^{208}Pb(n\gamma_0)$-reaction cross section 
$\sigma_{n\gamma_0}^{exp}$ \cite{3} in the GDR region
has been obtained in
\cite{17} with the use of interaction (1a) and the 
experimental cross section of photoabsorption $\sigma_{a}^{exp}$ \cite{20}.
The branching ratio $b^{(\gamma)}=\sum_c \sigma_{\gamma n_c}/\sigma_{a}
\simeq 10 \% $ calculated for the GDR energy is in qualitative agreement
with the value $b^{exp} \simeq 3-5 \%$ deduced from the 
$^{208}Pb(^{17}O,^{17}O' n_c)$-reaction cross-sections \cite{6}. 

In the present work we also use the separable p-h interaction~: \\
\\
$  F_{\tau}({\bf r, r'})=K_1{\bf r r'},$ \hfill (4) \\
\\
where the strength parameter $K_1=0.014$\ MeV$\cdot\mbox{fm}^{-2}$
is found so that the calculated mean dws energy 
would be equal to the experimental GDR energy 
 \cite{20}. The calculated cross section  
$\sigma_{n \gamma_0}$ is markedly underestimated in the GDR region
(by a factor of 3) as compared to the relevant experimental value.

The calculations performed with the use of interactions (1a) and
(4) show that the root-mean-squared dispersion of the dws energies
is somewhat less than the GDR spreading
width. For this reason one can roughly describe the cross sections
$\sigma_{\gamma n_c}$ in the GDR region by means
of the Breit-Wigner formulae in terms of some effective partial
neutron widths of the GDR $\Gamma_c^{(\gamma)}$
\cite{17}.
However, due to the GDR gross-structure, these widths can be
different, for example, from the effective partial 
neutron widths $\Gamma_c^{(n)}$
describing the GDR in the energy-averaged cross section
of the elastic neutron-nucleus scattering \cite{15}. This 
statement is illustrated by the calculation results given in 
Table~2.

4. The approach for description of the direct nucleon
decay of the giant resonances with not-too-large
excitation energy is briefly outlined. The approach is based on the exact
version of the continuum-RPA and phenomenological consideration of the 
doorway-state coupling to many-particle configurations. 
The approach is applied to describe the direct nucleon
decay of a number of GR in $^{208}Pb$ parent
nucleus.
The comparison of calculation
results with the relevant values deduced from experimental cross sections
shows that the use of the 
Landau-Migdal p-h interaction is preferable for the description of the isovector
giant resonances (IAR, GTR, GDR)
as compared with the use of separable interactions. In the case of the
GMR the situation is opposite.
The main conclusion follows from the above consideration:
the experimental and theoretical studies of the
direct nucleon decay of giant resonances 
with not-too-large excitation energy
allow one to get information on the particle-hole structure of these GR
and on the particle-hole interaction in nuclei.
For this reason continuation of these studies 
seems to be necessary.

5. The research described in this publication was made possible
in part by Grant MQ2000 from the International Science Foundation (ISF),
Grant MQ2300 from the ISF
and Russian Government, and Grant  95-02-05917-a from Russian Found of
Fundamental Researches (RFFR).
Two authors (G.A.Ch. and M.H.U.) are grateful to the 
International Soros Science Education Program for support
(respectively, Grants a73-f and 444p from the Open Society Institute, N.Y.).
One of the authors (M.H.U.) is grateful to the  
RFFR for travel-grant 95-02-07892d. 
Two authors (S.E.M. and M.H.U.) are grateful to Prof. M.N.Harakeh
as the Conference Chairperson for the local support.

\end{document}